\documentclass[orivec,conference]{llncs}
\newcommand{\jd}{\mbox{$\mi{JoinDseqs}$}\xspace}
\newcommand{\Jd}{\mbox{$\mi{JoinDseqs}^+$}\xspace}

\newcommand{\cdi}{\mbox{$\mi{DCDS}$}\xspace}
\newcommand{\Cdi}{\mbox{$\mi{DCDS^+}$}\xspace}
\newcommand{\PS}{\mbox{$\Psi$}\xspace}
\newcommand{\FI}{\mbox{$\Phi$}\xspace}
\newcommand{\olds}[4]{\mbox{$(\prob{#1}{#2},\pnt{#3})~\rightarrow #4$}}

\newcommand{\noolds}[4]{$(\prob{#1}{#2},#3)~\rightarrow #4$}
\newcommand{\Ods}[5]{\mbox{$(\prob{#1}{#2},\pnt{#3},#4)~\rightarrow #5$}}

\newcommand{\ods}[3]{\mbox{$(\pnt{#1},#2)~\rightarrow #3$}}
\newcommand{\nods}[3]{$(\pnt{#1},#2)~\rightarrow #3$}
\newcommand{\oDs}[3]{\mbox{$(#1,#2)~\rightarrow #3$}}
\newcommand{\noDs}[3]{$(#1,#2)~\rightarrow #3$}
\newcommand{\oods}[2]{\mbox{\pnt{#1}~$\rightarrow #2$}}
\newcommand{\ooods}[2]{\mbox{$#1 \rightarrow #2$}}

\newenvironment{mmproof}{\hspace{8pt}\ti{Proof:}}{}
\newcommand{\ie}{i.e., }

\newcommand{\pnt}[1]{{\mbox{$\vec{#1}$}}}
\newcommand{\ppnt}[2]{{\mbox{$\vec{#1}_{#2}$}}}

\newcommand{\bm}[1]{{\mbox{\boldmath $#1$}}}
\newcommand{\cof}[2]{\mbox{$#1_{\vec{#2}}$}}

\newcommand{\V}[1]{\mbox{$\mathit{Vars}(#1)$}}
\newcommand{\Va}[1]{\mbox{$\mathit{Vars}(\vec{#1})$}}

\newcommand{\s}[1]{\mbox{$\{#1\}$}}

\newcommand{\nGz}[2]{$G_{non-\{z\}}$}

\newcommand{\prr}[1]{\mi{Prev}(\boldsymbol{q})}

\newcommand{\mi}[1]{\mathit{#1}}
\newcommand{\ti}[1]{\textit{#1}}
\newcommand{\tb}[1]{\textbf{#1}}

\newcommand{\ttt}{\>\>\>}

\newcommand{\Tt}{\>\>}

\newcommand{\Prob}[2]{$\exists{#1} [#2]$}
\newcommand{\prob}[2]{\mbox{$\exists{#1} [#2]$}}

\newcommand{\DS}{\mbox{$\Omega$}\xspace}
\newcommand{\ecnf}{\ensuremath{\exists\mathrm{CNF}}\xspace}
\newcommand{\Comment}[1]{}

\usepackage[cmex10]{amsmath}
\usepackage{amsfonts}
\usepackage{appendix}
\usepackage{wrapfig}
\usepackage{graphicx}
\usepackage{enumerate}
\usepackage{multirow}
\usepackage{xspace}
\usepackage{cancel}
\begin{document}

\title{Quantifier Elimination With Structural Learning}


\author{Eugene Goldberg}
\institute{\email{eu.goldberg@gmail.com}}

\maketitle

\begin{abstract}
  We consider the Quantifier Elimination (QE) problem for
  propositional CNF formulas with existential quantifiers.  QE plays a
  key role in formal verification. Earlier, we presented an approach
  based on the following observation. To perform QE, one just needs to
  add a set of clauses depending on free variables that makes the
  quantified clauses (i.e. clauses with quantified variables)
  redundant.  To implement this approach, we introduced a branching
  algorithm making quantified clauses redundant in subspaces and
  merging the results of branches. To implement this algorithm we
  developed the machinery of D-sequents. A D-sequent is a record
  stating that a quantified clause is redundant in a specified
  subspace. Redundancy of a clause is a structural property (i.e. it
  holds only for a \ti{subset} of logically equivalent formulas as
  opposed to a \ti{semantic} property). So, re-using D-sequents is not
  as easy as re-using conflict clauses in SAT-solving. In this paper,
  we address this problem. We introduce a new definition of D-sequents
  that enables their re-usability. We develop a theory showing under
  what conditions a D-sequent can be safely re-used.
\end{abstract}

\section{Introduction}
Many verification problems can be cast as an instance of the
Quantifier Elimination (QE) problem or its
variations\footnote{In~\cite{hvc-14}, we introduced Partial QE (PQE) where only
a \ti{part} of the formula is taken out of the scope of
quantifiers. The appeal of PQE is twofold. First, many verification
problems like equivalence and model checking require partial rather
than complete QE~\cite{fmcad16,mc_no_inv}. Second, PQE is much simpler
to solve than QE. Both QE and PQE benefit from the results of this
paper. However, since QE is conceptually simpler than PQE, we picked
the former to introduce our new approach to D-sequent re-using.

}. So any progress in solving the
QE problem is of great importance. In this paper, we consider the QE
problem for propositional CNF formulas with existential quantifiers.
Given formula \prob{X}{F(X,Y)} where $X$ and $Y$ are sets of
variables, the QE problem is to find a quantifier-free formula
$F^*(Y)$ such that $F^* \equiv
\prob{X}{F}$. In~\cite{fmcad12,fmcad13}, we introduced a new approach
to QE based on the following observation. Let us call a
clause\footnote{A clause is a disjunction of literals. So, a CNF
  formula is a conjunction of clauses.} of $F$ an \bm{X}\tb{-clause}
if it contains at least one variable of $X$. Solving the QE problem
\prob{X}{F(X,Y)} reduces to finding formula $F^*(Y)$ implied by $F$
that makes the $X$-clauses of $F$ \ti{redundant} in \prob{X}{F^*
  \wedge F} (and so $F^* \equiv \prob{X}{F}$ holds). 

To implement the approach above, we introduced an algorithm called
\cdi (Derivation of Clause D-Sequents). \cdi is based on the following
three ideas. First, \cdi branches on variables of $F$ to reach a
subspace where proving redundancy\footnote{An $X$-clause $C$ is said to be redundant in \prob{X}{F} if
$\prob{X}{F} \equiv \prob{X}{F \setminus \s{C}}$. In this paper, we
use the standard convention of viewing a set of clauses
\s{C_1,\dots,C_n} as an alternative way to specify the CNF formula
$C_1 \wedge \dots \wedge C_n$. So, the expression $H \setminus \s{C}$
denotes the CNF formula obtained from $H$ by removing clause $C$.

} of
$X$-clauses (or making them redundant by adding a new clause) is
easy. Second, once an $X$-clause is proved redundant, \cdi stores this
fact in the form of a Dependency Sequent (\tb{D-sequent}). A D-sequent
is a record \olds{X}{F}{q}{C} where $C$ is an $X$-clause of $F$ and
\pnt{q} is an assignment to variables of $F$. This record states that
$C$ is redundant in \prob{X}{F} in subspace \pnt{q}. The third idea of
\cdi is to use a resolution-like operation called \ti{join} to merge
the results of branches. This join operation is applied to D-sequents
\olds{X}{F}{q'}{C} and \olds{X}{F}{q''}{C} derived in branches $v=0$
and $v=1$ where $v$ is a variable of $F$.  The result of this
operation is a D-sequent \olds{X}{F}{q}{C} where \pnt{q} does not contain
variable $v$.

To make \cdi more efficient, it is natural to try to re-use a
D-sequent \olds{X}{F}{q}{C} in every subspace \pnt{r} where $\pnt{q}
\subseteq \pnt{r}$ (i.e. \pnt{r} contains all the assignments of
\pnt{q}). However, here one faces the following problem. The
definition of D-sequent \olds{X}{F}{q}{C} implies that $C$ is also
redundant in subspace \pnt{q} for formulas \prob{X}{G} logically
equvialent to \prob{X}{F} where $G$ is a subset of $F$. However, this
may not be true for \ti{some} formulas \prob{X}{G}. Here is a simple
example of that. Let formula \prob{X}{F} contain two identical
$X$-clauses $C'$ and $C''$. Then D-sequents
\noolds{X}{F}{\emptyset}{C'} and \noolds{X}{F}{\emptyset}{C''}
hold. They state that $C'$ and $C''$ are redundant in \prob{X}{F}
\ti{individually}. However, in general, one cannot\footnote{That is
  $\prob{X}{F} \not\equiv \prob{X}{F \setminus \s{C',C''}}$.} drop
\ti{both} $C'$ and $C''$ from \prob{X}{F}. This means that, say, $C''$
may not be redundant in \prob{X}{F \setminus \s{C'}} despite the fact
that $F \setminus \s{C'} \equiv F$.

The problem above prevents \cdi from reusing D-sequents. The reason
why D-sequents cannot be re-used as easily as, say, conflict clauses
in SAT-solvers~\cite{grasp,chaff} is as follows. Redundancy of a
clause in a formula is a \tb{structural} property\footnote{This is
  true regardless of whether this formula has quantifiers.}. That is
the fact that clause $C$ is redundant in formula $F$ may not hold in a
formula $F'$ logically equivalent to $F$. On the other hand, re-using
a conflict clause $C$ is based on the fact that $C$ is implied by the
initial formula $F$ and implication is a \tb{semantic} property. That
is $C$ is implied by \ti{every} formula $F'$ logically equivalent to
$F$.

In this paper, we address the problem of re-usability of
D-sequents. Our approach is based on the following
observation. Consider the example above with two identical $X$-clauses
$C',C''$. The D-sequent \noolds{X}{F}{\emptyset}{C'} requires the
presence of clause $C'' \in F$. This means that $C''$ is supposed to
be proved redundant \ti{after} $C'$. On the contrary, the D-sequent
\noolds{X}{F}{\emptyset}{C''} requires the presence of $C' \in F$ and
hence $C''$ is proved redundant \ti{before} $C'$. So these D-sequents
have a conflict in the order of proving redundancy of $C'$ and $C''$.

To be able to identify order conflicts, we modify the definition of
D-sequents given in~\cite{fmcad13}.  A new D-sequent $S$ is a record
\Ods{X}{F}{q}{H}{C} where $H$ is a subset of $F \setminus \s{C}$.  This
D-sequent states that the clause $C$ is redundant in subspace \pnt{q}
in every formula \prob{X}{W} where $\prob{X}{W} \equiv \prob{X}{F}$ in
subspace \pnt{q} and $(H \cup \s{C}) \subseteq W \subseteq F$. Note
that if an $X$-clause of $H$ is proved redundant and removed from $F$,
the D-sequent $S$ is not applicable. So one can view $H$ as an
\tb{order constraint} stating that $S$ applies only if $X$-clauses of
$H$ are proved redundant \ti{after} $C$. In other words, one can
safely reuse $S$ in subspace \pnt{r} where $\pnt{q} \subseteq
\pnt{r}$, if none of the $X$-clauses of $H$ is proved redundant yet.

The contribution of this paper is as follows. First, we give the
necessary definitions and propositions explaining the semantics of
D-sequents stating redundancy of
$X$-clauses\footnote{In~\cite{fmcad13}, we just give basic definitions. In~\cite{fmsd14},
we do provide a detailed theoretical consideration but it is meant for
D-sequents introduced in~\cite{fmcad12} expressing redundancy of
variables rather than clauses. 
}
(Sections~\ref{sec:basic},~\ref{sec:bnd_pnts},~\ref{sec:div_conq}
and~\ref{sec:virt_red}.)  Second, we give a new definition of
D-sequents facilitating their re-using (Section~\ref{sec:dseqs}).  We
also introduce the notion of consistent D-sequents and show that they
can be re-used.  Third, we re-visit definitions of atomic D-sequents
(i.e. D-sequents stating trivial cases of redundancy) and the join
operation to accommodate the D-sequents of the new kind
(Sections~\ref{sec:atom_dseqs} and~\ref{sec:join_dseqs}).  Fourth, we
present \Cdi, a version of \cdi that can safely re-use D-sequents
(Section~\ref{sec:dcds_plus}).

\section{Basic Definitions}
\label{sec:basic}

In this paper, we consider only propositional CNF formulas. In the
sequel, when we say ``formula'' without mentioning quantifiers we mean
a quantifier-free CNF formula.  We consider \ti{true} and \ti{false}
as a special kind of clauses. A non-empty clause $C$ becomes \ti{true}
when it is satisfied by an assignment \pnt{q} \,i.e. when a literal of
$C$ is set to \ti{true} by \pnt{q}. A clause $C$ becomes \ti{false}
when it is falsified by \pnt{q} i.e.  when all the literals of $C$ are
set to \ti{false} by \pnt{q}.
%
%
\begin{definition}
 \label{def:ecnf}
  Let $F$ be a CNF formula and $X$ be a subset of variables of $F$. We
  will refer to formula \prob{X}{F} as \bm{\ecnf}.
\end{definition}
%
%
\begin{definition}
  \label{def:vars}
Let \pnt{q} be an assignment and $F$ be a CNF formula. \bm{\Va{q}}
denotes the variables assigned in \pnt{q}; \bm{\V{F}} denotes the set
of variables of $F$; \bm{\V{\prob{X}{F}}} denotes $\V{F} \setminus X$.
\end{definition}
%
%
\begin{definition}
  \label{def:cofactor}
Let $C$ be a clause, $H$ be a formula that may have quantifiers, and
\pnt{p} be an assignment. \bm{\cof{C}{p}} is true if $C$ is satisfied
by \pnt{p}; otherwise it is the clause obtained from $C$ by removing
all literals falsified by \pnt{p}. \bm{\cof{H}{p}} denotes the formula
obtained from $H$ by replacing $C$ with \cof{C}{p}.
\end{definition}
%
%
\begin{definition}
\label{def:formula-equiv}
Let $G, H$ be formulas that may have quantifiers. We say that $G, H$
are \tb{equivalent}, written \bm{G \equiv H}, if for all assignments
$\pnt{q}$ such that $\Va{q}$ $\supseteq (\V{G} \cup \V{H})$, we have
$\cof{G}{q} = \cof{H}{q}$.
\end{definition}
%
%
\begin{definition}
 \label{def:qe-solution}
The \tb{Quantifier Elimination (QE)} problem for \ecnf formula
\prob{X}{F(X,Y)} is to find a formula $F^*(Y)$ such that \bm{F^*
  \equiv \prob{X}{F}}.
\end{definition}

%
%
\begin{remark}
\label{rem:X_Y_sets}
From now on, we will use $Y$ and $X$ to denote sets of free and
quantified variables respectively. We will assume that variables
denoted by $x_i$ and $y_i$ are in $X$ and $Y$ respectively. When we
use $Y$ and $X$ in a quantifier-free formula we mean that, in the
context of QE, the set $X$ specifies the quantified variables.
\end{remark}
%
%
\begin{definition}
\label{def:X_clause}
 A clause $C$ of $F$ is called a \bm{Z}\tb{-clause} 
 if \V{C} $\cap~Z~\neq~\emptyset$. Denote by
           {\boldmath $F^{Z}$} the set of all $Z$-clauses of
           $F$.
\end{definition}
%
%
\begin{definition}
\label{def:red_cls}
Let $F$ be a CNF formula and $G \subseteq F$ (i.e. G is a non-empty
subset of clauses of $F$).  The clauses of $G$ are \textbf{redundant}
in $F$ if $F \equiv (F \setminus G)$.  The clauses of $G$ are
\textbf{redundant} in formula \prob{X}{F} if $\prob{X}{F} \equiv
\prob{X}{F \setminus G}$.
\end{definition}

Note that $F \equiv (F \setminus G)$ implies $\prob{X}{F} \equiv
\prob{X}{F \setminus G}$ but the opposite is not true.

\section{Clause Redundancy And Boundary Points}
\label{sec:bnd_pnts}
In this section, we explain the semantics of QE in terms so-called
boundary points.
%
%
\begin{definition}
\label{def:pnt}
Given assignment \pnt{p} and a formula $F$, we say that \pnt{p} is a
\textbf{point} of $F$ if $\V{F} \subseteq \Va{p}$.
\end{definition}

In the sequel, by ``assignment'' we mean a possibly partial one. To
refer to a \ti{full} assignment we will use the term ``point''.

%
%
\begin{definition}
  \label{def:bnd_pnt}
 Let $F$ be a formula and $Z \subseteq \V{F}$.  A point \pnt{p} of $F$
 is called a \bm{Z}\tb{-boundary point} of $F$ if a) $Z \neq
 \emptyset$ and b) $\cof{F}{p} = \mi{false}$ and c) every clause of
 $F$ falsified by \pnt{p} is a $Z$-clause and d) the previous
 condition breaks for every proper subset of $Z$.
\end{definition}

\begin{remark}
  Let $F(X,Y)$ be a CNF formula where sets $X$ and $Y$ are interpreted
  as described in Remark~\ref{rem:X_Y_sets}. In the context of QE, we
  will deal exclusively with $Z$-boundary points that falsify only
  $X$-clauses of $F$ and so $Z \subseteq X$ holds.
\end{remark}
%
%
\begin{example}
\label{exmp:bnd_pnt}
Let $X = \s{x_1,x_2,x_3}$ and $Y=\s{y_1,y_2}$. Let $F(X,Y)$ be a CNF
formula of four clauses: $C_1 = x_1 \vee x_2$, $C_2 = \overline{x}_1
\vee y_1$, $C_3 = x_1 \vee \overline{x}_3 \vee y_2$, $C_4 =
\overline{x}_2 \vee y_2$. The clauses of $F$ falsified by \pnt{p} =
$(x_1=0,x_2=0,x_3=1,y_1=0,y_2=0)$ are $C_1$ and $C_3$. One can verify
that \pnt{p} and the set $Z= \s{x_1}$ satisfy the four conditions of
Definition~\ref{def:bnd_pnt}, which makes \pnt{p} a \s{x_1}-boundary
point. The set $Z$ above is not unique. One can easily check that
\pnt{p} is also a \s{x_2,x_3}-boundary point.
\end{example}
The term ``boundary'' is justified as follows.  Let $F$ be a
satisfiable CNF formula with at least one clause.  Then there always
exists a \s{v}-boundary point of $F$, $v \in \V{F}$ that is different
from a satisfying assignment only in value of $v$.

%
%
\begin{definition}
\label{def:rem_bnd_pnt}
Given a CNF formula $F(X,Y)$ and a $Z$-boundary point \pnt{p} of $F$:
\begin{itemize} 
\item \pnt{p} is $X'$-\tb{removable} in $F$ if 1) $Z \subseteq X'
  \subseteq X$; and 2) there is a clause $C$ such that a) $F
  \Rightarrow C$; b) $\V{C} \cap X' = \emptyset$; and c)
  $\cof{C}{p} = \mathit{false}$.

\item \pnt{p} is
\tb{removable} in $\prob{X}{F}$ if \pnt{p} is $X$-removable in $F$.
\end{itemize} 
\end{definition}

In the above definition, notice that \pnt{p} is not a $Z$-boundary
point of $F \wedge C$ because \pnt{p} falsifies $C$ and $\V{C}\cap Z =
\emptyset$. So adding clause $C$ to $F$ eliminates \pnt{p} as a
$Z$-boundary point.
%
%
\begin{example}
Let us consider the \s{x_1}-boundary point \pnt{p} =
$(x_1=0,x_2=0,x_3=1,y_1=0,y_2=0)$ of Example~\ref{exmp:bnd_pnt}. Let
$C$ denote clause $C = y_1 \vee y_2$ obtained\footnote{See
  Definition~\ref{def:resol} of the resolution operation.} by
resolving $C_1$, $C_2$ and $C_4$ on variables $x_1$ and $x_2$.  Note
that set $Z = \s{x_1}$ and $C$ satisfy the conditions a),b) and c) of
Definition~\ref{def:rem_bnd_pnt} for $X' = X$.  So \pnt{p}~~is an
$X$-removable \s{x_1}-boundary point. After adding $C$ to $F$, \pnt{p}
is not an \s{x_1}-boundary point any more. Let us consider the point
\pnt{q}=$(x_1=0,x_2=0,x_3=1,y_1=1,y_2=1)$ obtained from \pnt{p} by
flipping values of $y_1$ and $y_2$. Both \pnt{p} and \pnt{q} have the
same set of falsified clauses consisting of $C_1$ and $C_3$. So, like
\pnt{p}, point \pnt{q} is an \s{x_1}-boundary point. However, no
clause $C$ implied by $F$ and consisting only of variables of $Y$ is
falsified by \pnt{q}. So, the latter, is an \s{x_1}-boundary point
that is \ti{not} $X$-removable.
\end{example}
%
%
\begin{proposition}
\label{prop:rem_bnd_pnt}
A $Z$-boundary point \pnt{p} of $F(X,Y)$ is removable in \prob{X}{F}, iff
one cannot turn \pnt{p} into an assignment satisfying $F$ by changing
only the values of variables of $X$.
\end{proposition}

The proofs are given in the appendix.
%
%
\begin{proposition}
\label{prop:red_cls_set}
Let $F(X,Y)$ be a CNF formula where $F^X \neq \emptyset$ (see
Definition~\ref{def:X_clause}). Let $G$ be a non-empty subset of
$F^X$.  The set $G$ is not redundant in \prob{X}{F} iff there is a
$Z$-boundary point \pnt{p} of $F$ such that a) every clause falsified
by \pnt{p} is in $G$ and b) \pnt{p} is $X$-removable in $F$.
\end{proposition}

Proposition~\ref{prop:red_cls_set} justifies the following strategy of
solving the QE problem. Add to $F$ a set $G$ of clauses that a) are
implied by $F$; b) eliminate every $X$-removable boundary point
falsifying a subset of $X$-clauses of $F$. By dropping all $X$-clauses
of $F$, one produces a solution to the QE problem.

\section{Quantifier Elimination By Branching} 
\label{sec:div_conq}
In this section, we explain the semantics of QE algorithm called
\cdi~\cite{fmcad13} (Derivation of Clause D-Sequents). A high-level
description of \cdi is given in Section~\ref{sec:dcds_recall}. \cdi is
a branching algorithm. Given a formula \prob{X}{F}, \cdi branches on
variables of $F$ until it proves that every $X$-clause is redundant in
the current subspace. (In case of a conflict, proving $X$-clauses of
$F$ redundant, in general, requires adding to $F$ a conflict clause.)
Then \cdi merges the results obtained in different branches to prove
that the $X$-clauses are redundant in the entire search space.  Below
we give propositions justifying the strategy of
\cdi. Proposition~\ref{prop:glb_to_loc} shows how to perform
elimination of removable boundary points of $F$ in the subspace
specified by assignment \pnt{q}. This is done by using formula
\cof{F}{q}, a ``local version'' of $F$.
Proposition~\ref{prop:elim_incr} justifies proving redundancy of
$X$-clauses of \cof{F}{q} incrementally.

%
%
Let \pnt{q} and \pnt{r} be assignments to a set of variables $Z$.
Since \pnt{q} and \pnt{r} are sets of value assignments to individual
variables of $Z$ one can apply set operations to them. We will denote
by $\pnt{r} \subseteq \pnt{q}$ the fact that \pnt{q} contains all
value assignments of \pnt{r}.  The assignment consisting of value
assignments of \pnt{q} and \pnt{r} is represented as $\pnt{q} \cup
\pnt{r}$.

%
%
\begin{proposition}
\label{prop:glb_to_loc}
Let \prob{X}{F(X,Y)} be an \ecnf and \pnt{q} be an assignment to
\V{F}. Let \pnt{p} be a $Z$-boundary point of $F$ where $\pnt{q}
\subseteq \pnt{p}$ and $Z \subseteq X$. Then if \pnt{p} is removable
in \prob{X}{F} it is also removable in \prob{X}{\cof{F}{q}}.
\end{proposition}
%
%
\begin{remark}
\label{rem:no_reverse}
One cannot reverse Proposition~\ref{prop:glb_to_loc}: a boundary point
may be $X$-removable in \cof{F}{q} and not $X$-removable in $F$.  For
instance, if $X=\V{F}$, a $Z$-boundary point \pnt{p} of $F$ where
$|Z|=1$ is removed from \prob{X}{F} only by adding an empty clause to
$F$. So if $F$ is satisfiable, \pnt{p} is not removable. Yet \pnt{p}
may be removable in \prob{X}{\cof{F}{q}} if \cof{F}{q} is
unsatisfiable. A ramification of the fact that
Proposition~\ref{prop:glb_to_loc} is not reversible is discussed in
Section~\ref{sec:virt_red}.

\end{remark}

%
%
\begin{proposition}
\label{prop:elim_incr}
Let \prob{X}{F(X,Y)} be an \ecnf and $H \subset F^X$ be redundant in
\prob{X}{F}.  Let an $X$-clause $C$ of $F \setminus H$ be redundant in
\prob{X}{F \setminus H}. Then $H \cup \s{C}$ is redundant in
\prob{X}{F}.
\end{proposition}

%
%
\begin{remark}
  \label{rem:red_subsp}
  To simplify the notation, we will sometimes use the expression
  ``clause $C$ is redundant in \prob{X}{F} in subspace \pnt{q}~''
  instead of saying ``clause \cof{C}{q} is redundant in
  \prob{X}{\cof{F}{q}}''.
\end{remark}

Proposition~\ref{prop:elim_incr} shows that one can prove redundancy
of, say, a set of $X$-clauses \s{C',C''} in \prob{X}{F} in subspace
\pnt{q} \ti{incrementally}.  This can be done by a) proving redundancy
of $C'$ in \prob{X}{F} in subspace \pnt{q} and c) proving redundancy
of $C''$ in formula $\prob{X}{F \setminus \s{C'} }$ in subspace
\pnt{q}.

\section{Virtual redundancy}
\label{sec:virt_red}

If a boundary point \pnt{p} is $X$-removable in \prob{X}{\cof{F}{q}},
this does not mean that it is $X$-removable in \prob{X}{F} (see
Remark~\ref{rem:no_reverse}). This fact leads to the following
problem.  Let \pnt{q} and \pnt{r} be two assignments to \V{F} and
$\pnt{q} \subset \pnt{r}$.  Suppose that clause $C$ is redundant in
\prob{X}{F} in subspace \pnt{q}. It is natural to expect that this
also holds in the smaller subspace \pnt{r}. However,
$\prob{X}{\cof{F}{q}} \equiv \prob{X}{\cof{F}{q} \setminus
  \cof{C}{q}}$ does not imply $\prob{X}{\cof{F}{r}} \equiv
\prob{X}{\cof{F}{r} \setminus \cof{C}{r}}$. In particular, due to this
problem, one cannot define the join operation in terms of redundancy
specified by Definition~\ref{def:red_cls}. To address this issue we
introduce the notion of \ti{virtual} redundancy.
%
%
\begin{definition}
\label{def:virt_red}
Let \prob{X}{F(X,Y)} be an \ecnf formula, \pnt{q} be an assignment
to \V{F}, and \cof{C}{q} be an $X$-clause of \cof{F}{q}. Let $B$ be
the set of points of $F$ such every $\pnt{p} \in B$ falsifies only
clause $C$ and is $X$-removable. Clause \cof{C}{q} is called
\tb{virtually redundant} in \prob{X}{\cof{F}{q}} if one of the two
conditions are true.
\begin{enumerate}
\item $B = \emptyset$ or
\item For every $\pnt{p} \in B$, there is an assignment \pnt{r} where
  $\pnt{q^*} \subseteq \pnt{r} \subset \pnt{q}$ such that \pnt{p} is
  \ti{not} $X$-removable in \cof{F}{r}. Here \pnt{q^*} is obtained
  from \pnt{q} by removing all value assignments to variables of $X$.
\end{enumerate}
\end{definition}

The first condition just means that $\prob{X}{\cof{F}{q} \setminus
  \cof{C}{q}} \equiv \prob{X}{\cof{F}{q}}$.  We will refer to this
type of redundancy (earlier specified by Definition~\ref{def:red_cls})
as \ti{regular} redundancy. Regular redundancy is a special case of
virtual redundancy.

%
%
\begin{proposition}
\label{prop:subspace}
Let \pnt{q} be an assignment to \V{F} and clause \cof{C}{q} be
redundant in \prob{X}{\cof{F}{q}}. Then, for every \pnt{r} such that
$\pnt{q} \subset \pnt{r}$, clause \cof{C}{r} is virtually redundant in
\prob{X}{\cof{F}{r}}.
\end{proposition}

From now on, when we say that a clause \cof{C}{r} is redundant in
\prob{X}{\cof{F}{r}} we mean that it is at least \ti{virtually
  redundant}. Note that, in general, proving virtual redundancy of $C$
in subspace \pnt{r} can be extremely hard. We avoid this problem by
using the notion of virtual of redundancy only if we have \ti{already
  proved} that $C$ is redundant in a subspace containing subspace
\pnt{r}. (For instance, we have already proved that $C$ is redundant
in \prob{X}{F} in subspace \pnt{q} where $\pnt{q} \subset \pnt{r}$.)

\section{Dependency Sequents (D-sequents)}
\label{sec:dseqs}
In this section, we give a new definition of D-sequents that is
different from that of~\cite{fmcad13}.

%
%
\begin{definition}
\label{def:dseq}
Let \prob{X}{F} be an \ecnf formula. Let \pnt{q} be an assignment to
\V{F} and $C \in F^X$ and $H \subseteq (F \setminus \s{C})$. A
dependency sequent (\tb{D-sequent}) $S$ has the form
\Ods{X}{F}{q}{H}{C}. It states that clause \cof{C}{q} is redundant in
every formula \prob{X}{\cof{W}{q}} logically equivalent to
\prob{X}{\cof{F}{q}} where \mbox{$H \cup \s{C} \subseteq W \subseteq
  F$}.  The assignment \pnt{q} and formula $H$ are called the
\tb{conditional} and \tb{order constraint} of $S$ respectively. We
will refer to $W$ as a \tb{member formula} for $S$.
\end{definition}

Definition~\ref{def:dseq} implies that the D-sequent $S$ becomes
inapplicable if a clause of $H$ is removed from $F$. So, $S$ is meant
to be used in situations where the $X$-clauses of $H$ are proved
redundant \ti{after} $C$ (hence the name ``order constraint'').  As we
mentioned in the introduction, in~\cite{fmcad13}, a D-sequent implies
redundancy of clause $C$ in \prob{X}{F} and in (some) logically
equivalent formulas \prob{X}{W} where $W \subseteq F$. In
Definition~\ref{def:dseq}, the set of formulas \prob{X}{W} where $C$
is redundant in subspace \pnt{q} is specified \ti{precisely}.  We will
say that a D-sequent \Ods{X}{F}{q}{H}{C} is \tb{fragile} if $H$
contains at least one $X$-clause.  Such a D-sequent becomes
inapplicable if an $X$-clause of $H$ is proved redundant before
$C$. If $H$ does not contain $X$-clauses, the D-sequent above is
called \tb{robust}. A robust D-sequent is not affected by the order in
which $X$-clauses are proved redundant.

%
%
\begin{remark}
\label{rem:short_dseqs}
We will abbreviate D-sequent \Ods{X}{F}{q}{H}{C} to \ods{q}{H}{C} if
formula \prob{X}{F} is known from the context. We will further reduce
\ods{q}{H}{C} to \oods{q}{C} if $H = \emptyset$ i.e. if no order
constraint is imposed.
\end{remark}

There are two ways to produce D-sequents. First, one can generate an
``atomic'' D-sequent that states a trivial case of redundancy.  The
three atomic types of D-sequents are presented in
Section~\ref{sec:atom_dseqs}. Second, one can use a pair of existing
D-sequents to generate a new one by applying a resolution-like
operation called \ti{join} (Section~\ref{sec:join_dseqs}).

\section{Atomic D-sequents}
\label{sec:atom_dseqs}
In this section we describe D-sequents called atomic. These D-sequents
are generated when redundancy of a clause can be trivially
proved. Similarly to~\cite{fmcad13}, we introduce atomic D-sequents of
three kinds. However, in contrast to~\cite{fmcad13}, we consider
D-sequents specified by Definition~\ref{def:dseq}. In particular, we
show that D-sequents of the first kind are robust whereas D-sequents
of the second and third kind are fragile.
%
%
\subsection{Atomic D-sequents of the first kind}
\begin{proposition}
  \label{prop:sat_cls}
 Let \prob{X}{F} be an \ecnf and $C \in F$ and $v \in \V{C}$. Let
 assignment $v=b$ where $b \in \s{0,1}$ satisfy $C$. Then D-sequent
 \ooods{(v=b)}{C} holds. We will refer to it as an atomic D-sequent of
 the \tb{first kind}.
\end{proposition}
%
%
\begin{example}
  Let \prob{X}{F} be an \ecnf and $C=\overline{x}_1 \vee y_5$ be a
  clause of $F$.  Since $C$ is satisfied by assignments $x_1 = 0$ and
  $y_5=1$, D-sequents \ooods{(x_1=0)}{C} and \ooods{(y_5=1)}{C} hold.
\end{example}
%
%
\subsection{Atomic D-sequents of the second kind}
\label{ssec:second_kind}
\begin{proposition}
  \label{prop:uns_cls}
 Let \prob{X}{F} be an \ecnf formula and \pnt{q} be an assignment to
 \V{F}.  Let $B,C$ be two clauses of $F$. Let \cof{C}{q} be an
 $X$-clause and \cof{B}{q} imply \cof{C}{q} (i.e. every literal of
 \cof{B}{q} is in \cof{C}{q}). Then the D-sequent \ods{q}{H}{C} holds
 where $H = \s{B}$.  We will refer to it as an atomic D-sequent of the
 \tb{second kind}.
\end{proposition}
\begin{example}
  Let \prob{X}{F} be an \ecnf formula.  Let $B=y_1 \vee x_2$ and $C =
  x_2 \vee \overline{x}_3$ be $X$-clauses of $F$. Let
  $\pnt{q}=(y_1=0)$.  Since \cof{B}{q} implies \cof{C}{q}, the
  D-sequent \ods{q}{\s{B}}{C} holds. Since \cof{B}{q} \,is an
  $X$-clause, this D-sequent is fragile.
\end{example}

%
%
\subsection{Atomic D-sequents of the third kind}
\label{ssec:third_kind}
To introduce atomic D-sequents of the third kind, we need to make a
few definitions.
%
%
\begin{definition}
\label{def:resol}
Let $C'$ and $C''$ be  clauses having opposite
  literals of exactly one variable $v \in \V{C'} \cap \V{C''}$. 
The clause $C$ consisting of all literals of $C'$ and $C''$ but
those of $v$ is called the \tb{resolvent} of $C'$,$C''$ on $v$.
Clause $C$ is said to be obtained by \tb{resolution} on $v$.
Clauses $C'$,$C''$ are called \tb{resolvable} on $v$.
\end{definition}
%
%
\begin{definition}
\label{def:blk_cls}
 A clause $C$ of a CNF formula $F$ is called \tb{blocked} at variable
 $v$, if no clause of $F$ is resolvable with $C$ on $v$.  The notion
 of blocked clauses was introduced in~\cite{blocked_clause}.
\end{definition}

%
%
\begin{proposition}
  \label{prop:dseq_third_kind}
Let \prob{X}{F} be an \ecnf formula. Let $C$ be an $X$-clause of $F$
and $v \in\V{C} \cap X$. Let $C_1,\dots,C_k$ be the clauses of $F$
that can be resolved with $C$ on variable $v$. Let
\noDs{\ppnt{q}{1}}{H_1}{C_1},$\dots$,\noDs{\ppnt{q}{k}}{H_k}{C_k} be a
consistent set\footnote{We will introduce the notion of a consistent set of D-sequents later,
see Definition~\ref{def:cons_dseqs}. Consistency of D-sequents in
Proposition~\ref{prop:dseq_third_kind} means that $C_1,\dots,C_k$ are
redundant together in subspace
\pnt{q}=$\bigcup\limits_{i=1}^{i=k}\ppnt{q}{i}$. So clause $C$ is
blocked at variable $v$ in subspace \pnt{q}.
} of D-sequents.  Then
D-sequent \noDs{q}{H}{C} holds where
\pnt{q}=$\bigcup\limits_{i=1}^{i=k}\ppnt{q}{i}$ and $H =
\bigcup\limits_{i=1}^{i=k}H_i$. We will refer to it as an atomic
D-sequent of the \tb{third kind}.
\end{proposition}

Note that, in general, a D-sequent of the third kind is fragile.

%
%
\begin{example}
  Let \prob{X}{F(X,Y)} be an \ecnf formula. Let $C_3,C_6,C_8$ be the
  only clauses of $F$ with variable $x_5 \in X$ where $C_3 = x_5
  \wedge x_{10} $, $C_6 = \overline{x}_5 \wedge y_1$, $C_8 =
  \overline{x_5} \vee y_3 \vee y_5$. Note that assignment $y_1=1$
  satisfies clause $C_6$.  So the D-sequent \ooods{(y_1=1)}{C_6}
  holds. Suppose that D-sequent \noDs{\pnt{r}}{\s{C_{10}}}{C_8} holds
  where $C_{10}$ is a clause of $F$ and $\pnt{r} = (y_2=0,x_{10}=1)$.
  From Proposition~\ref{prop:dseq_third_kind} it follows that
  D-sequent \noDs{\pnt{q}}{\s{C_{10}}}{C_3} holds where $\pnt{q} =
  (y_1=1,y_2=0,x_{10}=1)$.
\end{example}

\section{Join Operation}
\label{sec:join_dseqs}
In this section, we describe the operation of joining D-sequents that
produces a new D-sequent from two parent D-sequents. In contrast
to~\cite{fmcad13}, the join operation introduced here is applied to
D-sequents \ti{with order constraints}.

%
%
\begin{definition}
\label{def:res_part_assgns}
Let \pnt{q'} and \pnt{q''} be assignments in which exactly one
variable $v \in \Va{q'} \cap \Va{q''}$ is assigned different values.
The assignment \pnt{q} consisting of all the value assignments of
\pnt{q'} and \pnt{q''} but those to $v$ is called the \tb{resolvent}
of \pnt{q'},\pnt{q''} on $v$.  Assignments \pnt{q'},\pnt{q''} are
called \tb{resolvable} on $v$.
\end{definition}
%
%
\begin{proposition}
\label{prop:join_rule}
Let \prob{X}{F} be an \ecnf formula for which D-sequents
\nods{q'}{H'}{C} and \ods{q''}{H''}{C} hold. Let \pnt{q'}, \pnt{q''}
be resolvable on $v \in \V{F}$ and \pnt{q} be the resolvent of
\pnt{q'} and \pnt{q''}.  Let $H = H' \cup H''$.  Then the D-sequent
\ods{q}{H}{C} holds.
\end{proposition}

%
%
\begin{definition}
\label{def:join_rule}
We will say that the D-sequent \ods{q}{H}{C} of
Proposition~\ref{prop:join_rule} is produced by \tb{joining
  D-sequents} \ods{q'}{H'}{C} and \ods{q''}{H''}{C} \tb{at} \bm{v}.
\end{definition}
%
%
\begin{remark}
  \label{rem:ord_constr}
  Note that the D-sequent $S$ produced by the join operation has a
  \ti{stronger} order constraint than its parent D-sequents. The
  latter have order constraints $H'$ and $H''$ in subspaces $v=0$ and
  $v=1$, whereas $S$ has the same order constraint $H = H' \cup H''$
  in either subspace. Due to this ``imprecision'' of the join
  operation, a set of D-sequents with conflicting order constraints
  can still be correct (see Section~\ref{sec:reuse} and
  Subsection~\ref{ssec:dcds_corr}).
\end{remark}

\section{Re-usability of D-sequents}
\label{sec:reuse}
To address the problem of D-sequent re-using, we introduce the notion
of composability. Informally, a set of D-sequents is composable if the
clauses stated redundant individually are also redundant
\ti{collectively}.  Robust D-sequents are always composable. So they
can be re-used in any context like conflict clauses in
SAT-solvers. However, this is not true for fragile D-sequents. Below,
we show that such D-sequents are composable if they are consistent.
So it is safe to re-use a fragile D-sequent in a subspace \pnt{q}, if
it is consistent with the D-sequents already used in subspace \pnt{q}.

%
%
\begin{definition}
  Assignments \pnt{q'} and \pnt{q''} are called \tb{compatible} if
  every variable from $\Va{q'} \cap \Va{q''}$ is assigned the same
  value.
\end{definition}

%
%
\begin{definition}
\label{def:comp_dseqs}
  Let \,\prob{X}{F} be an \,\ecnf.  A set of D-sequents
  \oDs{\ppnt{q}{1}}{H_1}{C_1},$\dots$, \oDs{\ppnt{q}{k}}{H_k}{C_k} is
  called \tb{composable} if the clauses \s{C_1,\dots,C_k} are
  redundant collectively as well. That is $\prob{X}{F} \equiv$
  \Prob{X}{F \setminus \s{C_1,\dots,C_k}} holds in subspace \pnt{q}
  where \pnt{q}=$\bigcup\limits_{i=1}^{i=k}\ppnt{q}{i}$.
\end{definition}

%
%
\begin{definition}
\label{def:cons_dseqs}
  Let \prob{X}{F} be an \ecnf. A set of D-sequents
  \oDs{\ppnt{q}{1}}{H_1}{C_1},$\dots$, \oDs{\ppnt{q}{k}}{H_k}{C_k} is
  called \tb{consistent} if
  \begin{itemize}
  \item every pair of assignments \pnt{q_i},\pnt{q_j}, $1 \leq i,j \leq k$
    is compatible;
  \item there is a total order $\pi$ over clauses of
    $\bigcup\limits_{i=1}^{i=k} H_i \cup \s{C_i}$ that \vspace{4pt}
    satisfies the order constraints of these D-sequents
    i.e. $\forall{C} \in H_i$, $\pi(C_i) < \pi(C)$ holds where
    $i=1,\dots,k$.
  \end{itemize} 
\end{definition}

%
%
\begin{proposition}
\label{prop:cons_dseqs}
 Let \prob{X}{F} be an \ecnf. Let \oDs{\ppnt{q}{1}}{H_1}{C_1},$\dots$,
 $(\ppnt{q}{k},H_k)$ $\rightarrow C_k$ be a consistent set of
 D-sequents. Then these D-sequents are composable and hence clauses
 \s{C_1,\dots,C_k} are collectively redundant in \prob{X}{F} in
 subspace \pnt{q} where \pnt{q}=$\bigcup\limits_{i=1}^{i=k}
 \ppnt{q}{i}$.
\end{proposition}
%
%
\begin{remark}
  The fact that D-sequents $S_1,\dots,S_k$ are inconsistent does not
  necessarily mean that these D-sequents are not composable.  As we
  mentioned in Remark~\ref{rem:ord_constr}, as far as order
  constraints are concerned, the join operation is not ``precise''.
  This means that if the D-sequents above are obtained by applying the
  join operation, their order-inconsistency may be artificial. An
  example of that is the QE procedure called \cdi~\cite{fmcad13}. As
  we explain in Subsection~\ref{ssec:dcds_corr}, if one uses the new
  definition of D-sequents (i.e.  Definition~\ref{def:dseq}), the
  D-sequents produced by \cdi are, in general, inconsistent.  However,
  \cdi is provably correct~\cite{fmsd14}.
\end{remark}
%
%
\begin{remark}
  Let \prob{X}{F(X,Y)} be an \ecnf and $R(X,Y)$ be the set of clauses
  added to $F$ by a QE-solver. Let $F^X \cup R^X =\s{C_1,\dots,C_k}$
  (i.e. the latter is the set of all $X$-clauses of $F \cup R$).  This
  QE-solver terminates when the set\footnote{This set consists of the
    clauses of $R$ that depend \ti{only} on variables of $Y$.} $R
  \setminus R^X$ is sufficient to derive consistent D-sequents
  $(\prob{X}{F \wedge R}$,$\emptyset,H_1)$ $\rightarrow C_1$, $\dots$,
  $(\prob{X}{F \wedge R}$,$\emptyset,H_k)$ $\rightarrow C_k$. From
  Proposition~\ref{prop:cons_dseqs} it follows, that all $X$-clauses
  can be dropped from \prob{X}{F \wedge R}. The resulting formula
  $F^*(Y)$ consisting of clauses of $F \cup R \setminus
  \s{C_1,\dots,C_k}$ is logically equivalent to \prob{X}{F}.
\end{remark}

\section{Two Useful Transformations Of D-sequents}
\label{sec:trans_dseqs}
In this section, we describe two transformations that are useful for a
QE-solver based on the machinery of D-sequents. Since a QE-solver has
to add new clauses once in a while, D-sequents of different branches
are, in general, computed with respect to different formulas.  In
Subsection~\ref{ssec:align}, we describe a transformation meant for
``aligning'' such D-sequents. In Subsection~\ref{ssec:mak_rob}, we
describe a transformation meant for relaxing the order constraint of a
D-sequent. In Section~\ref{sec:dcds_plus}, this transformation is used
to generate a consistent set of D-sequents.

%
%
\subsection{D-sequent alignment}
\label{ssec:align}
According to Definition~\ref{def:dseq}, a D-sequent holds with respect
to a particular \ecnf formula \prob{X}{F}.
Proposition~\ref{prop:dseq_add_cls} shows that this D-sequent also
holds after adding to $F$ implied clauses.
%
%
\begin{proposition}
\label{prop:dseq_add_cls}
Let D-sequent \Ods{X}{F}{q}{H}{C} hold and $R$ be a CNF formula
implied by $F$. Then D-sequent \Ods{X}{F \wedge R}{q}{H}{C} holds too.
\end{proposition}

Proposition~\ref{prop:dseq_add_cls} is useful in aligning D-sequents
derived in different branches. Suppose that \Ods{X}{F}{q'}{H'}{C} is
derived in the current branch of the search tree where the last
assignment is $v=0$. Suppose that \Ods{X}{F \wedge R}{q''}{H''}{C} is
derived after flipping the value of $v$ from 0 to 1. Here $R$ is the
set of clauses implied by $F$ that has been added to $F$ before the
second D-sequent was derived.  One cannot apply the join operation to
these D-sequents because they are computed with respect to different
formulas. Proposition~\ref{prop:dseq_add_cls} allows one to replace
\Ods{X}{F}{q'}{H'}{C} with \Ods{X}{F\wedge R}{q'}{H'}{C}. The latter
can be joined with \Ods{X}{F\wedge R}{q''}{H''}{C} at variable $v$.

%
%
\subsection{Making a D-sequent more robust}
\label{ssec:mak_rob}
In this subsection, we give two propositions showing how one can make
a D-sequent $S$ more robust. Proposition~\ref{prop:subst} introduces a
transformation that removes a clause from the order constraint of $S$
possibly adding to the latter some other
clauses. Proposition~\ref{prop:mak_rob} describes a scenario where by
repeatedly applying this transformation one can remove a clause from
the order constraint of $S$ \ti{without} adding any other clauses.

%
%
\begin{proposition}
  \label{prop:subst}  
Let \prob{X}{F} be an \ecnf.  Let \oDs{\pnt{q'}}{H'}{C'} and
$(\pnt{q''},H'')$$\rightarrow C''$ be two D-sequents forming a
consistent set (see Definition~\ref{def:cons_dseqs}). Let $C''$ be in
$H'$. Then  D-sequent \ods{q}{H}{C'} holds where $\pnt{q} =
\pnt{q'} \cup \pnt{q''}$ and $H = (H' \setminus \s{C''}) \cup H''$.
\end{proposition}

%
%
\begin{proposition}
\label{prop:mak_rob}
Let \prob{X}{F} be an \ecnf and $(\ppnt{q}{1},H_1)$ $\rightarrow
C_1$,$\dots$, $(\ppnt{q}{k},H_k)$$\rightarrow C_k$ be consistent
D-sequents where $H_i \subseteq \s{C_1,\dots,C_k}$,
$i=1,\dots,k$. Assume, for the sake of simplicity, that the numbering
order is consistent with the order constraints. Let $C_m$ be in $H_i$.
Then, by repeatedly applying the transformation of
Proposition~\ref{prop:subst}, one can produce D-sequent \ods{q}{H_i
  \setminus \s{C_m}}{C_i} where $\ppnt{q}{i}~\subseteq~\pnt{q}
\subseteq \ppnt{q}{i} \cup \bigcup\limits_{j=m}^{j=k} \ppnt{q}{j}$.
\end{proposition}

\section{Recalling \cdi}
\label{sec:dcds_recall}
In~\cite{fmcad13}, we described a QE algorithm called \cdi (Derivation
of Clause D-Sequents) that did not re-use D-sequents. We Recall \cdi in
Subsections~\ref{ssec:dcds} and~\ref{ssec:dcds_corr}.  

%
%
\subsection{A brief description of \cdi}
\label{ssec:dcds}

The pseudocode of \cdi is given\footnote{For the sake of simplicity, Figure~\ref{fig:dcds} gives a very
abstract view of \cdi. For instance, we omit the lines of code where
new clauses are generated. Our objective here is just to show the part
of \cdi where D-sequents are involved. A more detailed description
of \cdi can be found in~\cite{fmcad13}.
} in
Fig.~\ref{fig:dcds}. \cdi uses the old definition of a D-sequent
lacking an order constraint. \cdi accepts three parameters: formula
\prob{X}{F} (denoted as \FI), the current assignment \pnt{q} \,and the
set of active D-sequents \DS. (If an $X$-clause of $F$ is proved
redundant in subspace \pnt{q}, this fact is stated by a D-sequent.
This D-sequent is called \tb{active}). \cdi returns the final formula
\prob{X}{F} (where $F$ consists of the initial clauses and derived
clauses implied by $F$) and the set \DS of current active D-sequents.
\DS has an active D-sequent for every $X$-clause of $F$. The
conditional of this D-sequent is a subset of \pnt{q}. In the first
call of \cdi, the initial formula \prob{X}{F} is used and \pnt{q} and
\DS are empty sets. A solution $F^*(Y)$ to the QE problem at hand is
obtained by dropping the $X$-clauses of the final formula \prob{X}{F}
and removing the quantifiers.

\cdi starts with examining the $X$-clauses whose redundancy is not
proved yet.  Namely, \cdi checks if the redundancy of such clauses can
be established by atomic D-sequents (line 1) introduced in
Section~\ref{sec:atom_dseqs}. If all $X$-clauses are proved redundant,
\cdi terminates returning the current formula \prob{X}{F} and the
current set of active D-sequents (lines 2-3). Otherwise, \cdi moves to
the branching part of the algorithm (lines 4-9).

%
%
%
\setlength{\intextsep}{2pt}
\setlength{\textfloatsep}{5pt}
\begin{wrapfigure}{L}{2.3in}
\small
\vspace{5pt}
// \FI denotes  \prob{X}{F} \\
// \pnt{q} is an assignment to \V{F} \\
// \DS~denotes the current set of active \\
//~~~~~D-sequents \\
\vspace{-10pt}
\begin{tabbing}
aaaa\=bb\= c\= dddddddddddd\= \kill
\cdi\!\!(\FI,\pnt{q},\DS)\{\\
\tb{\scriptsize{1}}\> $\DS := \DS \cup \mi{AtomDseqs}(\Phi,\pnt{q},\DS)$ \\
\tb{\scriptsize{2}}\>if ($\mi{EveryXclauseRedund}(\Phi,\DS)$) \\
\tb{\scriptsize{3}}\Tt return($\Phi,\DS$)\\
\> - - - - - - - - - - - -  \\
\tb{\scriptsize{4}}\> $v := \mi{PickVar}(F,\pnt{q},\DS)$ \\
\tb{\scriptsize{5}}\> $(\Phi,\DS_0):=$\cdi\!\!(\FI,$\pnt{q}\cup\s{v=0}$,\DS)\\
\tb{\scriptsize{6}}\> $\DS := \mi{DropInapplic}(\DS_0,v)$ \\
\tb{\scriptsize{7}}\> $(\Phi,\DS_1):=$\cdi\!\!(\FI,$\pnt{q} \cup\s{v=1}$,\DS)\\
\tb{\scriptsize{8}}\> $\DS := \mi{JoinDseqs}(\Phi,v,\DS_0,\DS_1)$\\
\tb{\scriptsize{9}}\> return($\Phi,\DS$)\} \\
\end{tabbing} 
\vspace{-15pt}
\caption{\cdi procedure}
\label{fig:dcds}
\end{wrapfigure}

First, \cdi picks a variable $v$ to branch on (line 4). Then it
explores the branch $\pnt{q} \cup \s{v = 0}$ (line 5). The set $\DS_0$
returned in this branch, in general, contains D-sequents whose
conditionals include assignment $v=0$. These D-sequents are
inapplicable in branch $v=1$ and so they are discarded (line 6). After
that \cdi explores branch $\pnt{q} \cup \s{v = 1}$ (line 7) returning
a set of D-sequents $\DS_1$.  Then \cdi generates a set of D-sequents
\DS whose conditionals do not depend on $v$ (line 8). Set \DS consists
of two parts. The first part comprises of the D-sequents of $\DS_0$
that do not depend on $v$. The second part consists of the D-sequents
obtained by joining D-sequents of $\DS_0$ and $\DS_1$ that do depend
on $v$. Finally, \cdi terminates returning \prob{X}{F} and \DS.

%
%
\subsection{Correctness of DCDS}
\label{ssec:dcds_corr}
As we mentioned earlier, \cdi employs D-sequents introduced
in~\cite{fmcad13} that lack order constraints.  A D-sequent $S$
of~\cite{fmcad13} states redundancy of a clause $C$ in formula
\prob{X}{F} in subspace \pnt{q}. Besides, clause $C$ is also assumed
to be redundant in (some) formulas \prob{X}{W} logically equivalent to
\prob{X}{F} where $W$ is a subset of $F$. The problem here is that the
set of formulas \prob{X}{W} for which $S$ guranatees redundancy of
clause $C$ in subspace \pnt{q} is not specified
precisely. Nevertheless, \cdi is provably
correct\footnote{In~\cite{fmsd14}, we proved the correctness of a
  similar algorithm. This proof applies to \cdi.}.

There are three reasons why \cdi is correct despite the fact that it
uses a ``sloppy'' definition of a D-sequent. First, \cdi does not
\ti{re-use} D-sequents. After generating a new D-sequent by the join
operation, \cdi discards the parent clauses of this D-sequent.
Second, in every branch of the search tree the $X$-clauses are proved
redundant in some order which makes them \ti{composable}. Third, by
joining composable D-sequents obtained in branch $v=0$ with composable
D-sequents obtained in branch $v=1$ one produces a set of composable
D-sequents.

\section{Introducing \Cdi}
\label{sec:dcds_plus}

In this section, we describe a modification of \cdi that re-uses
D-sequents. We will refer to it as \Cdi. The pseudocode of \Cdi is
shown in Fig.~\ref{fig:dcds+}. In comparison to \cdi, \Cdi has one
more input parameter: a set \PS of D-sequents stored to re-use. The
four lines where \Cdi behaves differently from \cdi are marked with an
asterisk.

%
%
%
\setlength{\intextsep}{2pt}
\setlength{\textfloatsep}{5pt}
\begin{wrapfigure}{l}{2.3in}
\small
\vspace{5pt}
// \FI denotes  \prob{X}{F} \\
// \pnt{q} is an assignment to \V{F} \\
// \DS~denotes the current set of active \\
//~~~~~D-sequents \\
// \PS denotes the set of stored D-sequents \\
\vspace{-10pt}
\begin{tabbing}
aaaa\=bb\= c\= dddddddddddd\= \kill
\Cdi\!\!(\FI,\pnt{q},\DS,\PS)\{\\
\tb{\scriptsize{1*}}\> $\DS := \DS \cup \mi{ReuseDseqs}(\FI,\pnt{q},\PS,\DS)$ \\
\tb{\scriptsize{2*}}\> $\DS := \DS \cup \mi{AtomDseqs}(\Phi,\pnt{q},\DS)$ \\
\tb{\scriptsize{3}}\>if ($\mi{EveryXclauseRedund}(\Phi,\DS)$) \\
\tb{\scriptsize{4}}\Tt return($\Phi,\DS,\PS$)\\
\> - - - - - - - - - - - -  \\
\tb{\scriptsize{5}}\> $v := \mi{PickVar}(F,\pnt{q},\DS)$ \\
\tb{\scriptsize{6}}\> $(\Phi,\DS_0,\PS):=$\Cdi\!\!($\Phi$,$\pnt{q}\cup\!(v=0)$,\DS,\PS)\\
\tb{\scriptsize{7}}\> $\DS := \mi{DropInapplic}(\DS_0,v)$ \\
\tb{\scriptsize{8}}\> $(\Phi,\DS_1,\PS):=$\Cdi\!\!($\Phi$,$\pnt{q}\cup\!(v=1)$,\DS,\PS)\\
\tb{\scriptsize{9*}}\> $\DS := \mi{JoinDseqs}^+(\Phi,v,\DS_0,\DS_1)$\\
\tb{\scriptsize{10*}}\> $\PS := \PS \cup DseqsToStore(\DS)$ \\
\tb{\scriptsize{11}}\> return($\Phi,\DS,\PS$)\} \\
\end{tabbing} 
\vspace{-15pt}
\caption{\Cdi procedure}
\vspace{5pt}
\label{fig:dcds+}
\end{wrapfigure}

The difference between \Cdi and \cdi is as follows.  First, \Cdi uses
the new definition of D-sequents and thus keeps track of order
constraints. (In particular, \Cdi stores order constraints when
generating the atomic D-sequents of the second/third kind.  For that
reason, line 2 is marked with an asterisk.) Second, \Cdi tries to
re-use D-sequents stored in \PS. Namely, if an $X$-clause $C$ is not
proved redundant yet, \Cdi checks if there is a D-sequent of \PS a)
that states redundancy of $C$; b) whose conditional \pnt{r} satisfies
$\pnt{r} \subseteq \pnt{q}$ and c) whose order constraint is
consistent with those of active D-sequents. Third, \Cdi stores some of
new D-sequents obtained by the join operation (line 10).

%
%
\setlength{\intextsep}{4pt}
\setlength{\textfloatsep}{10pt}
\begin{wrapfigure}{l}{2in}
\small
\begin{tabbing}
aaa\=bb\= cc\= ddddddd\= \kill
$\mi{JoinDseqs}^+(\Phi,v,\DS_0,\DS_1)$\{\\
\tb{\scriptsize{1}}\> $\DS := \mi{SymmDseqs}(\DS_0)$ \\
\tb{\scriptsize{2}}\> $G := \mi{FormXcls}(\DS_0 \setminus \DS)$ \\
\tb{\scriptsize{3}}\> foreach ($C \in G$) \{ \\
\tb{\scriptsize{4}}\Tt  $(S_0,S_1) := \mi{ExtrDseqs}(\DS_0,\DS_1,C)$\\
\tb{\scriptsize{5}}\Tt  $S := \mi{join}(S_0,S_1,v)$ \\
\tb{\scriptsize{6}}\Tt  if ($\mi{incons}(\DS \cup \s{S})$) \\
\tb{\scriptsize{7}}\ttt   $S := \mi{FixDseq}(S,\DS,\DS_0,\DS_1)$ \\
\tb{\scriptsize{8}}\Tt  $\DS := \DS \cup \s{S}$ \}\\
\tb{\scriptsize{9}}\> return(\DS);\} \\
\end{tabbing} 
\vspace{-20pt}
\caption{$\mi{JoinDseqs}^+$ procedure}
\label{fig:join_dseqs+}
\end{wrapfigure}

The final difference between \Cdi and \cdi is as follows.  The order
in which $X$-clauses are proved redundant in the two branches
generated by splitting on variable $v$ can be different. So if one
just joins D-sequents obtained in those branches at variable $v$ (as
it is done by the \jd procedure of \cdi), an inconsistent set of
D-sequents can be generated. If no D-sequents are re-used, as in \cdi,
this inconsistency does not mean that the D-sequents of this set are
not composable\footnote{As we mentioned in
  Remark~\ref{rem:ord_constr}, an inconsistency can be introduced by
  the imprecision of the join operation with respect to order
  constraints.}  (see Subsection~\ref{ssec:dcds_corr}). However,
re-using D-sequents, as it is done in \Cdi, may produce inconsistent
D-sequents that are \ti{indeed} not composable. For that reason, in
\Cdi, a modification of \jd called \Jd is used.

The pseudocode of \Jd is shown in Fig.~\ref{fig:join_dseqs+}. The
objective of \Jd is to generate a set of \ti{consistent} D-sequents
that do not depend on variable $v$. The resulting D-sequents are
accumulated in \DS. \Jd starts by initializing \DS with the D-sequents
that are already symmetric in $v$ i.e. their conditionals do not
contain an assignment to variable $v$.  Then \Jd forms the set $G$ of
$X$-clauses whose D-sequents are asymmetric in $v$.

The main part of \Jd consists of a loop (lines 3-8) where, for every
clause $C$ of $G$, a D-sequent whose conditional is symmetric in $v$
is built. First, \Jd extracts D-sequents $S_0$ and $S_1$ of clause $C$
and joins them at variable $v$ to produce a D-sequent $S$ (lines
4-5). If the D-sequents of \DS become inconsistent after adding $S$,
\Jd calls $\mi{FixDseq}$ to produce a D-sequent $S$ that preserves the
consistency of \DS.  Proposition~\ref{prop:mak_rob} shows that it is
always possible.  Namely, one can always relax the order constraints
of $S_0$ and $S_1$ thus relaxing that of $S$.  In particular, one can
totally eliminate order constraints of $S_0$ and $S_1$, which makes
them (and hence $S$) robust.

\bibliographystyle{plain}
\bibliography{short_sat}
\appendix
\setcounter{proposition}{0}
\vspace{10pt}
\noindent\tb{\large{Appendix}} \\
\vspace{10pt}

The appendix contains proofs of the propositions listed in the paper.
We also give proofs of lemmas used in the proofs of propositions.

\section*{Propositions of
Section~\ref{sec:bnd_pnts}: Clause Redundancy And Boundary Points}

%
%

\begin{proposition}
A $Z$-boundary point \pnt{p} of $F(X,Y)$ is removable in \prob{X}{F}, iff
one cannot turn \pnt{p} into an assignment satisfying $F$ by changing
only the values of variables of $X$.
\end{proposition}
\begin{mmproof} \ti{If part}. Assume the contrary. That is \pnt{p} \,is
not removable while no satisfying assignment can be obtained from
\pnt{p} by changing only values of variables of $X$. Let $C$ be the
clause consisting of all variables of $Y$ and falsified by
\pnt{p}. Since \pnt{p} is not removable, clause $C$ is not implied by
$F$. This means that there is an assignment \pnt{s} that falsifies $C$
and satisfies $F$. By construction, \pnt{s} and \pnt{p} \,have identical
assignments to variables of $Y$. Thus, \pnt{s} can be obtained from
\pnt{p} by changing only values of variables of $X$ and we have a
contradiction.
\vspace{3pt}

\noindent\textit{Only if part.} Assume the contrary. That is \pnt{p}
is removable but one can obtain an assignment \pnt{s} satisfying $F$
from \pnt{p} \,by changing only values of variables of $X$.  Since
\pnt{p} \,is removable, there is a clause $C$ that is implied by $F$
and falsified by \pnt{p} and that depends only of variables of
$Y$. Since \pnt{s} and \pnt{p} \,have identical assignments to
variables of $Y$, point \pnt{s} \,falsifies $C$. However, since
\pnt{s} satisfies $F$, this means that $C$ is not implied by $F$ and
we have a contradiction. $\square$
\end{mmproof}

%
%
\begin{proposition}
Let $F(X,Y)$ be a CNF formula where $F^X \neq \emptyset$ (see
Definition~\ref{def:X_clause}). Let $G$ be a non-empty subset of
$F^X$.  The set $G$ is not redundant in \prob{X}{F} iff there is a
$Z$-boundary point \pnt{p} of $F$ such that a) every clause falsified
by \pnt{p} is in $G$ and b) \pnt{p} is $X$-removable in $F$.
\end{proposition}
%
%
\begin{mmproof}
Let $H$ denote $F \setminus G$. Given a point \pnt{p}, let
(\pnt{x},\pnt{y}) specify the assignments of \pnt{p} to the variables
of $X$ and $Y$ respectively. \\
\noindent\textit{If part.} Assume the contrary, \ie 
 there is an $X$-removable point \pnt{p}=(\pnt{x},\pnt{y}) of $F$ but
 $G$ is redundant in \prob{X}{F} and so
 $\prob{X}{F} \equiv \prob{X}{H}$.  Since \pnt{p} \,is a boundary point,
 $F(\pnt{p}) = 0$. Since
\pnt{p} \,is removable, $\cof{(\prob{X}{F})}{y} = 0$. On the other hand,
since \pnt{p} \,falsifies only clauses of $G$ it satisfies $H$.
Hence $\cof{(\prob{X}{H})}{y} = 1$ and
$\cof{(\prob{X}{F})}{y} \neq \cof{(\prob{X}{H})}{y}$, which leads to a
contradiction.

\vspace{3pt}
\noindent\textit{Only if part.}  Assume the contrary, \ie set $G$ is
not redundant (and hence $\prob{X}{F} \not\equiv \prob{X}{H}$) and
there does not exist an $X$-removable $Z$-boundary point of $F$
falsifying only clauses of $G$.  Let \pnt{y} be an assignment to $Y$
such that $\cof{(\prob{X}{F})}{y} \neq \cof{(\prob{X}{H})}{y}$.
Consider the following two cases.
\begin{itemize}
\item   $\cof{(\prob{X}{F})}{y}=1$ and $\cof{(\prob{X}{H})}{y}=0$.
Then there exists an assignment \pnt{x} to $X$ such that
(\pnt{x},\pnt{y}) satisfies $F$. Since every clause of $H$ is in $F$,
formula $H$ is also satisfied by \pnt{p}. So we have a contradiction.
\item $\cof{(\prob{X}{F})}{y}=0$ and $\cof{(\prob{X}{H})}{y}=1$. Then
  there exists an assignment \pnt{x} to variables of $X$ such that
  (\pnt{x},\pnt{y}) satisfies $H$.  Since $\cof{F}{y} \equiv 0$, point
  (\pnt{x},\pnt{y}) falsifies $F$. Since $H(\pnt{p})=1$,
  (\pnt{x},\pnt{y}) is a $Z$-boundary point of $F$ that falsifies only
  clauses of $G$.  Since $\cof{F}{y} \equiv 0$, (\pnt{x},\pnt{y}) is
  an $X$-removable $Z$-boundary point of $F$, which leads to a
  contradiction.~$\square$
\end{itemize}
\end{mmproof}

\section*{Propositions of Section~\ref{sec:div_conq}: Quantifier Elimination By Branching}
%
%
\begin{proposition}
Let \prob{X}{F(X,Y)} be an \ecnf and \pnt{q} be an assignment to
\V{F}. Let \pnt{p} be a $Z$-boundary point of $F$ where $\pnt{q}
\subseteq \pnt{p}$ and $Z \subseteq X$. Then if \pnt{p} is removable
in \prob{X}{F} it is also removable in \prob{X}{\cof{F}{q}}.
\end{proposition}
%
%
\begin{mmproof}
Assume the contrary. That is \pnt{p} is removable in $\prob{X}{F}$ but
is not removable in \prob{X}{\cof{F}{q}}. The fact that \pnt{p} is
removable in \prob{X}{F} means that there is a clause $C$ implied by
$F$ and falsified by \pnt{p} that consists only of variables of $Y$.
Since \pnt{p} is not removable in \prob{X}{\cof{F}{q}}, from
Proposition~\ref{prop:rem_bnd_pnt} it follows that an assignment
\pnt{s} satisfying \cof{F}{q} can be obtained from \pnt{p} by changing
only values of variables of $X \setminus \Va{q}$.  By construction,
\pnt{p} and \pnt{s} have identical assignments to variables of $Y$. So
\pnt{s} has to falsify $C$. On the other hand, by construction,
$\pnt{q} \subseteq \pnt{s}$.  So, the fact that \pnt{s} satisfies
\cof{F}{q} implies that \pnt{s} satisfies $F$ too. Since \pnt{s}
falsifies $C$ and satisfies $F$, clause $C$ is not implied by $F$ and
we have a contradiction. $\square$
\end{mmproof}

%
%
\begin{proposition}
Let \prob{X}{F(X,Y)} be an \ecnf and $H \subset F^X$ be redundant in
\prob{X}{F}.  Let an $X$-clause $C$ of $F \setminus H$ be redundant in
\prob{X}{F \setminus H}. Then $H \cup \s{C}$ is redundant in
\prob{X}{F}.
\end{proposition}
%
%
\begin{mmproof}
  Denote $H \cup \s{C}$ as $H'$.  Assume the contrary, i.e.  $H'$ is
  not redundant in \prob{X}{F}. Then, from
  Proposition~\ref{prop:red_cls_set}, it follows that $F$ has an
  $X$-removable boundary \pnt{p} \,such that every clause falsified by
  \pnt{p} \,is in $H'$. Denote as $H''$ the subset of clauses of $H'$
  falsified by \pnt{p}.  Let us consider the two possible cases.
  \begin{itemize}
  \item Clause $C$ is not in $H''$. In this case, $H''$ is a subset of
    $H$ and the existence of \pnt{p} means that $H$ is not
    redundant in \prob{X}{F}. So we have a contradiction.
  \item Clause $C$ is in $H''$. Redundancy of $C$ in \prob{X}{F
    \setminus H} means that one can turn \pnt{p} \,into an assignment
    \pnt{s} satisfying $F \setminus H$ by flipping values of variables
    from $X$. Since \pnt{p} \,is $X$-removable, \pnt{s} \,falsifies
    $F$. The only clauses of $F$ falsified by \pnt{s} are those of
    $H$. Since \pnt{p} and \pnt{s} \,have identical values assigned to
    $Y$, point \pnt{s} \,is $X$-removable as well.  Then the existence
    of \pnt{s} \,means that $H$ is not redundant in \prob{X}{F} and we
    have a contradiction.  $\square$
  \end{itemize}

\end{mmproof}

\section*{Propositions of Section~\ref{sec:virt_red}: Virtual Redundancy}
%
%
\begin{proposition}
Let \pnt{q} be an assignment to \V{F} and clause \cof{C}{q} be
redundant in \prob{X}{\cof{F}{q}}. Then, for every \pnt{r} such that
$\pnt{q} \subset \pnt{r}$, clause \cof{C}{r} is virtually redundant in
\prob{X}{\cof{F}{r}}.
\end{proposition}
\begin{mmproof}
  Let a point \pnt{p} of $F$ falsify only \cof{C}{r} and be
  $X$-removable.  Then \pnt{p} falsifies only clause \cof{C}{q} of
  \cof{F}{q}.  Since \cof{C}{q} is redundant in \prob{X}{\cof{F}{q}},
  point \pnt{p} \,is not $X$-removable in \cof{F}{q}. $\square$
\end{mmproof}

\section*{Propositions of Section~\ref{sec:atom_dseqs}: Atomic D-sequents}
%
%
\begin{proposition}
  Let \prob{X}{F} be an \ecnf. Let $v \in \V{C}$ and by assigning
  value $b$ to $v$ where $b \in \s{0,1}$ one satisfies $C$. Then
  D-sequent \ooods{(v=b)}{C} holds We will refer to it as a
  \tb{D-sequent of the first kind}.
\end{proposition}  
\begin{mmproof}
  Let \pnt{q} denote $(v=b)$. Let $W$ be a member formula for the
  D-sequent \ooods{(v=b)}{C}. Hence, $W$ contains $C$. Clause
  \cof{C}{q} is \ti{true} and so is redundant in \cof{W}{q} and hence,
  in \prob{X}{\cof{W}{q}}.  $\square$
\end{mmproof}

%
%
\begin{proposition}
  Let \prob{X}{F} be an \ecnf formula and \pnt{q} be an assignment to
  \V{F}.  Let $B,C$ be two clauses of $F$. Let \cof{C}{q} be an
  $X$-clause and \cof{B}{q} imply \cof{C}{q} (i.e. every literal of
  \cof{B}{q} is in \cof{C}{q}). Then the D-sequent \ods{q}{H}{C} holds
  where $H = \s{B}$ if \cof{B}{q} is an $X$-clause and $H = \emptyset$
  otherwise.  We will refer to the D-sequent above as an atomic
  D-sequent of the \tb{second kind}.
\end{proposition}
\begin{mmproof}
 Let $W$ be a member formula for the D-sequent \ods{q}{H}{C}. Then
 $\s{B,C} \subseteq W$. Indeed, if \cof{B}{q} is an $X$-clause then $B
 \in H$ and $H \subseteq W$. Otherwise, $B$ is in $W$ because $W$ is
 obtained from $F$ by removing only clauses that are $X$-clauses in
 subspace \pnt{q}.  Since \cof{B}{q} implies \cof{C}{q}, then
 $\cof{W}{q} \equiv \cof{W}{q} \setminus \s{\cof{C}{q}}$.$\square$
\end{mmproof}

%
%
\begin{lemma}
\label{lem:blk_cls}
Let \prob{X}{F} be an \ecnf formula and \pnt{q} be an assignment to
\V{F}. Let $C$ be an $X$-clause of $F$ not satisfied by \pnt{q} and $v
\in X$ be a variable of $C$ such that $v \not\in \Va{q}$.  Let clause
\cof{C}{q} be blocked at $v$ in \cof{F}{q}.  Then \cof{C}{q} is
redundant in \prob{X}{\cof{F}{q}}.
\end{lemma}
\begin{mmproof}
Assume the contrary i.e. \cof{C}{q} is not redundant in
\prob{X}{\cof{F}{q}}. Then there is a $Z$-boundary point \pnt{p} where
$Z \subseteq X$ that falsifies only \cof{C}{q} and is $X$-removable in
\cof{F}{q}. Let \pnt{p'} be the point obtained from \pnt{p} by
flipping the value of $v$. Consider the following two possibilities.
\begin{itemize}
\item \pnt{p'} satisfies \cof{F}{q}. Then \pnt{p} is not
 $X$-removable and we have a contradiction.
\item \pnt{p'} falsifies a clause \cof{C'}{q} of \cof{F}{q}. Then
  \cof{C}{q} and \cof{C'}{q} are resolvable on variable $v$ and we
  have a contradiction again. $\square$
\end{itemize}
\end{mmproof}

%
%
\begin{proposition}
Let \prob{X}{F} be an \ecnf formula. Let $C$ be an $X$-clause of $F$
and $v \in\V{C} \cap X$. Let $C_1,\dots,C_k$ be the clauses of $F$
that can be resolved with $C$ on variable $v$. Let
\noDs{\ppnt{q}{1}}{H_1}{C_1},$\dots$,\noDs{\ppnt{q}{k}}{H_k}{C_k} be a
consistent set of D-sequents.  Then D-sequent \noDs{q}{H}{C} holds
where \pnt{q}=$\bigcup\limits_{i=1}^{i=k}\ppnt{q}{i}$ and $H =
\bigcup\limits_{i=1}^{i=k}H_i$. We will refer to it as an atomic
D-sequent of the \tb{third kind}.
\end{proposition}
\begin{mmproof}
  Let $W$ be a member formula for the D-sequent \noDs{q}{H}{C}.  Since
  $H_i \subseteq H$ holds, $W$ is a member formula for D-sequent
  \noDs{\ppnt{q}{i}}{H_i}{C_i} too. So $C_i$ is redundant in
  \prob{X}{W} in subspace \ppnt{q}{i}.  Since $\ppnt{q}{i} \subseteq
  \pnt{q}$, from Proposition~\ref{prop:subspace} it follows, that
  $C_i$ is redundant in \prob{X}{W} in subspace \pnt{q} too. Since
  D-sequents \noDs{\ppnt{q}{1}}{H_1}{C_1}, $\dots$,
  \noDs{\ppnt{q}{k}}{H_k}{C_k} are consistent, the clauses
  $C_1,\dots,C_k$ are redundant together in \prob{X}{W} in subspace
  \pnt{q}. So clause $C$ is blocked in \prob{X}{W} at variable $v$ in
  subspace \pnt{q}. From Lemma~\ref{lem:blk_cls} it follows that $C$
  is redundant in \prob{X}{W} in subspace \pnt{q}.  $\square$
\end{mmproof}

\section*{Propositions of Section~\ref{sec:join_dseqs}: Join Operation}

%
%
\begin{proposition}
  Let \,\prob{X}{F} be an \ecnf formula. Let D-sequents \ods{q'}{H'}{C}
and \ods{q''}{H''}{C} hold. Let \pnt{q'}, \pnt{q''} be resolvable on
$v \in \V{F}$ and \pnt{q} be the resolvent of \pnt{q'} and \pnt{q''}.
Let $H = H' \cup H''$.  Then the D-sequent \ods{q}{H}{C} holds.
\end{proposition}
\begin{mmproof}
Denote by $S'$ and $S$ the D-sequents \ods{q'}{H'}{C} and
\ods{q}{H}{C} respectively. Assume that $S$ does not hold. Then there
is a member formula $W$ of $S$ such that $C$ is not redundant in
\prob{X}{\cof{W}{q}} even virtually.  This means that
\begin{itemize}
\item the set $B$ of Definition~\ref{def:virt_red} is not empty and
\item there is a point $\pnt{p} \in B$ that is $X$-removable in every
  formula \cof{W}{r} where $\pnt{q^*} \subseteq \pnt{r} \subset
  \pnt{q}$ (see Definition~\ref{def:virt_red}).
\end{itemize}
 Assume for the sake of clarity that \pnt{p} has the same assignment
 to $v$ as \pnt{q'}. Note that, since $H' \subseteq H$, $W$ is a
 member formula of the D-sequent \ods{q'}{H'}{C}. So, \cof{C}{q'} is
 redundant in \prob{X}{\cof{W}{q'}}.  Consider the following
 possibilities.
 \begin{itemize}
 \item Point \pnt{p} \,is not $X$-removable in \cof{W}{q'} and $v \in
   X$. Then it is not $X$-removable in \cof{W}{s} where \pnt{s} is
   obtained from \pnt{q'} by dropping the assignment to $v$. Since
   $\pnt{s} \subset \pnt{q}$ \,holds, this contradicts the fact that
   \pnt{p} has to be $X$-removable in every subspace $\pnt{q^*}
   \subseteq \pnt{r} \subset \pnt{q}$.
 \item Point \pnt{p} \,is not $X$-removable in \cof{W}{q'} and $v
   \not\in X$.  Then \pnt{p} \,is not $X$-removable in \cof{W}{q}.
   This contradicts the fact that $\pnt{p} \in B$.
   \item Point \pnt{p} \,is $X$-removable in \cof{W}{q'} and it is
     $X$-removable in \cof{W}{r} for every $\vec{q'}^* \subseteq
     \pnt{r} \subset \pnt{q'}$.  Then \cof{C}{q'} is not redundant in
     \prob{X}{\cof{W}{q'}}, \vspace{3pt} which contradicts the fact
     that D-sequent $S'$ holds.
 \item Point \pnt{p} is $X$-removable in \cof{W}{q'}, and $v \in X$
   and \pnt{p} is not $X$-removable in \cof{W}{r} where $\vec{q'}^*
   \subseteq \pnt{r} \subset \pnt{q'}$.  Then it is not $X$-removable
   in \cof{W}{r^*} obtained from \pnt{r} by dropping the assignment to
   $v$, if any (regardless of whether or not $v$ is in $X$).  Since
   $\pnt{r^*} \subset \pnt{q}$ holds, this contradicts the fact that
   \pnt{p} \,has to be $X$-removable in every subspace $\pnt{q^*}
   \subseteq \pnt{r} \subset \pnt{q}$.  $\square$
 \end{itemize}

\end{mmproof}

\section*{Propositions of Section~\ref{sec:reuse}: Re-usability of D-sequents}
%

%
%
\begin{proposition}
Let \prob{X}{F} be an \ecnf. Let \oDs{\ppnt{q}{1}}{H_1}{C_1},$\dots$,
$(\ppnt{q}{k},H_k)$ $\rightarrow C_k$ be a consistent set of
D-sequents. Then these D-sequents are composable and hence clauses
\s{C_1,\dots,C_k} are collectively redundant in \prob{X}{F} in
subspace \pnt{q} where \pnt{q}=$\bigcup\limits_{i=1}^{i=k}
\ppnt{q}{i}$.
\end{proposition}
\begin{mmproof}
Since $\ppnt{q}{i} \subseteq \pnt{q}$, $i=1,\dots,k$, from
Proposition~\ref{prop:subspace} it follows that D-sequents
\ods{q}{H_i}{C_i}, $i=1,\dots,k$ hold. Assume for the sake of
simplicity that $\pi(C_i) < \pi(C_j)$ if $i < j$. Then one can prove
redundancy of clauses $C_i$, $i=1,..,k$ in subspace \pnt{q} \,in the
order they are numbered. Denote \ods{q}{H_i}{C_i} as $S_i$,
$i=1,\dots,k$.  Denote by $F_{i+1}$ formula $F \setminus
\s{C_1,\dots,C_i}$, $i=1,\dots,k$. Formula $F_1$ is set to $F$.  Since
$S_1$ is applicable to $F_1$ one can remove $C_1$ in subspace \pnt{q}
producing formula $F_2$.  Note that $F_2$ is a member formula for
$S_2$. By applying $S_2$ to $F_1$ one removes clause $C_2$ in subspace
\pnt{q} producing $F_3$, a member formula for $S_3$. Going on in such
a manner one eventually produces formula $F_{k+1}$ thus showing that
\s{C_1,\dots,C_k} are redundant in \prob{X}{F} in subspace
\pnt{q}. $\square$
\end{mmproof}

\section*{Propositions of
  Section~\ref{sec:trans_dseqs}: Two Useful Transformations Of D-sequents}
%
%
\begin{lemma}
\label{lem:mk_sat}
Let \prob{X}{F(X,Y)} be an \ecnf formula. Let \pnt{q} be an assignment
to \V{F}. Let $C$ be an $X$-clause redundant in \prob{X}{\cof{F}{q}}.
Let $C$ be also redundant in formula \prob{X}{\cof{W}{q}} where $W$ is
obtained from $F$ by dropping $X$-clauses that are redundant in
\prob{X}{\cof{F}{q}}. Then every point \pnt{p} falsifying only clause
\cof{C}{q} of \cof{W}{q} can be turned into a point satisfying $F$ by
changing values of (some) variables of $X$.
\end{lemma}
\begin{mmproof}
Denote by $C_1,\dots,C_k$ the $X$-clauses dropped from $F$ to obtain
$W$. Assume that these clauses were dropped in the numbering
order. That is clause $C_i$ is redundant in \prob{X}{W_i} in subspace
\pnt{q} where $W_1$ = $F$ and $W_{i+1} = W_i \setminus \s{C_i}$,
$i=1,\dots,k$. So $W_{k+1} = F \setminus \s{C_1,\dots,C_k} = W$.

Let \pnt{p} be a point falsifying only $C$ of $W$ in subspace
\pnt{q}. Since clause $C$ is redundant, one can turn \pnt{p}
into a point satisfying $W$ in subspace \pnt{q} by changing
values of variables of $X$. Denote the new point as \pnt{p}
again. If \pnt{p} satisfies $F$ then we are done. Otherwise,
\pnt{p} falsifies some clauses $C_1,\dots,C_k$.

Let $C_i$ be the clause with the largest index that is falsified by
\pnt{p}. Note that \pnt{p} falsifies only clause $C_i$ in $W_i$ in
subspace \pnt{q}. Since $C_i$ is redundant in \prob{X}{W_i} in
subspace \pnt{q}, one can turn \pnt{p} \,into an assignment satisfying
$W_i$ in subspace \pnt{q} \,by changing only assignments to $X$. Denote
the new point as \pnt{p} again. If \pnt{p} satisfies $F$ we are
done. Otherwise, \pnt{p} falsifies some clauses $C_1,\dots,C_{i-1}$.

Going on in such a manner one eventually builds a point satisfying $F$
that is obtained from the very first point \pnt{p} by changing only
assignments to variables of $X$.  $\square$
\end{mmproof}

%
%
\begin{proposition}
Let D-sequent \Ods{X}{F}{q}{H}{C} hold and $R$ be a CNF formula
implied by $F$. Then D-sequent \Ods{X}{F \wedge R}{q}{H}{C} holds too.
\end{proposition}
\begin{mmproof}
Let $W^R$ be a member formula for \Ods{X}{F \wedge R}{q}{H}{C}.
Denote by $W$ the formula $W^R \setminus R$. Note that $W$ is a member
formula for \Ods{X}{F}{q}{H}{C}. If $W = W^R$, then $C$ is redundant
in \prob{X}{W} (and hence in \prob{X}{W^R}) in subspace \pnt{q}.  Now
consider the case $W \subset W^R$. Assume that $C$ is not redundant in
\prob{X}{W^R} in subspace \pnt{q}. Then there is a point \pnt{p} such
that
\begin{itemize}
\item \pnt{p} falsifies only clause $C$ of $W_R$ in subspace \pnt{q}
  where $\pnt{q} \subseteq \pnt{p}$
\item \pnt{p} is $X$-removable for every \pnt{r} obtained from \pnt{q}
  \,by dropping (some) assignments to $X$.
\end{itemize}
Since $C$ is redundant in \prob{X}{W} in subspace \pnt{q}, from
Lemma~\ref{lem:mk_sat}, it follows that one can turn \pnt{p} into
point \pnt{p^*} satisfying $F$ by changing only values of variables of
$X$. Since $R$ is implied by $F$, \pnt{p^*} satisfies $F \wedge R$ as
well. Hence \pnt{p^*} satisfies $W_R$ and \pnt{p} is not $X$-removable
in subspace \pnt{r} obtained by dropping from \pnt{q} all assignments
to $X$. So we have a contradiction.  $\square$
\end{mmproof}
%
%
\begin{lemma}
\label{lem:two_red_cls}
Let \prob{X}{F} be an \ecnf formula and \pnt{q} be an assignment to
\V{F}. Let $C'$ and $C''$ be $X$-clauses of $F$ and \s{C',C''} be
redundant in \prob{X}{F} in subspace \pnt{q}. Then clause $C'$ is
redundant in \prob{X}{F \setminus \s{C''}} in subspace \pnt{q}.
\end{lemma}
\begin{mmproof}
Assume that $C'$ is not redundant in \prob{X}{F \setminus \s{C''}}.
in subspace \pnt{q}. Then there is a point \pnt{p} falsifying only
\cof{C'}{q} that is $X$-removable in every subspace $\pnt{q^*}
\subseteq \pnt{r} \subseteq \pnt{q}$ (see
Definition~\ref{def:virt_red}). Then, from
Proposition~\ref{prop:red_cls_set} it follows that clauses \s{C',C''}
are not redundant in \prob{X}{F} in subspace \pnt{q} (even virtually).  $\square$
\end{mmproof}

%
%
\begin{lemma}
  \label{lem:subst}
  Let \prob{X}{F} be an \ecnf. Let D-sequents \ods{q'}{H'}{C'} and
  $(\pnt{q''},H'')$ $\rightarrow C''$ hold where $\pnt{q''} \subseteq
  \pnt{q'}$ and $H'' \subset H'$ and $C'' \in H'$ and $C' \not\in
  H''$. Then the D-sequent \ods{q'}{H' \setminus \s{C''}}{C'} holds.
\end{lemma}
\begin{mmproof}
  Let $W$ be a member formula for \ods{q'}{H' \setminus
    \s{C''}}{C'}. Let us show that $C'$ is redundant in \prob{X}{W} in
  subspace \pnt{q'} and so \ods{q'}{H' \setminus \s{C''}}{C'}
  holds. Consider the following two situations.  First, assume that
  clause $C'' \in W$. Then $W$ is a member formula for the D-sequent
  \ods{q'}{H'}{C'} and hence $C'$ is redundant in \prob{X}{W} in
  subspace \pnt{q'}.

  Now assume that clause $C'' \not\in W$.  Denote D-sequents
  \ods{q'}{H'}{C'} and \ods{q'}{H''}{C''} (computed with respect to
  \prob{X}{F}) as $S'$ and $S''$ respectively. Denote formula $W \cup
  \s{C''}$ by $W''$. Note that $W''$ is a member formula for $S'$ and
  $S''$. Besides, $S'$ and $S''$ are consistent. So, in particular,
  $S''$ can be used after $S'$ . By applying $S'$ and $S''$ one shows
  that \s{C',C''} are redundant in \prob{X}{W''} in subspace
  \pnt{q'}. From Lemma~\ref{lem:two_red_cls} it follows that $C'$ is
  redundant in \prob{X}{W} in subspace \pnt{q'}.  $\square$
\end{mmproof}

%
%
\begin{lemma}
  \label{lem:stren_order}
  Let \prob{X}{F} be an \ecnf. Let D-sequent \ods{q}{H}{C} hold. Let
  $G$ be an arbitrary subset of $X$-clauses of $F \setminus H$. Then
  D-sequent \ods{q}{H \cup G}{C} holds too.
\end{lemma}
\begin{mmproof}
Let $W$ be a member formula for \ods{q}{H \cup G}{C}. Then $W$ is a
member formula for \ods{q}{H}{C}. So, $C$ is redundant in \prob{X}{W}
in subspace \pnt{q}.~$\square$
\end{mmproof}
%
%
\begin{proposition}
Let \prob{X}{F} be an \ecnf.  Let \oDs{\pnt{q'}}{H'}{C'} and
$(\pnt{q''},H'')$$\rightarrow C''$ be two D-sequents forming a
consistent set (see Definition~\ref{def:cons_dseqs}). Let $C''$ be in
$H'$. Then  D-sequent \ods{q}{H}{C'} holds where $\pnt{q} =
\pnt{q'} \cup \pnt{q''}$ and $H = (H' \setminus \s{C''}) \cup H''$.
\end{proposition}
\begin{mmproof}
  From Proposition~\ref{prop:subspace} it follows that D-sequent
  \ods{q}{H'}{C'} holds. From Lemma~\ref{lem:stren_order} it follows
  that \ods{q}{H' \cup H''}{C'} holds too. The latter and D-sequent
  \ods{q''}{H''}{C''} satisfy the conditions of
  Lemma~\ref{lem:subst}. (Note that $H''$ cannot contain $C'$ because
  \ods{q'}{H'}{C'} and $(\pnt{q''},H'')$ $\rightarrow C''$ are
  consistent.)  This entails that \ods{q}{H}{C'} holds.  $\square$
\end{mmproof}

%
%
\begin{proposition}
Let \prob{X}{F} be an \ecnf and $(\ppnt{q}{1},H_1)$ $\rightarrow
C_1$,$\dots$, $(\ppnt{q}{k},H_k)$$\rightarrow C_k$ be consistent
D-sequents where $H_i \subseteq \s{C_1,\dots,C_k}$,
$i=1,\dots,k$. Assume, for the sake of simplicity, that the numbering
order is consistent with the order constraints. Let $C_m$ be in $H_i$.
Then, by repeatedly applying the transformation of
Proposition~\ref{prop:subst}, one can produce D-sequent \ods{q}{H_i
  \setminus \s{C_m}}{C_i} where $\ppnt{q}{i}~\subseteq~\pnt{q}
\subseteq \ppnt{q}{i} \cup \bigcup\limits_{j=m}^{j=k} \ppnt{q}{j}$.
\end{proposition}
\begin{mmproof}
 Let $S_j$ denote D-sequent \oDs{\ppnt{q}{j}}{H_j}{C_j},
 $j=1,\dots,k$. Note that $i < m$ holds, otherwise, $S_m$ would be
 used before $S_i$ proving $C_m$ redundant and thus making $S_i$
 inapplicable.  Let us use Proposition~\ref{prop:subst} to remove
 clause $C_m$ from the order constraint of $S_m$. This produces a new
 D-sequent $S$ equal to \nods{q}{(H_i \cup H_m) \setminus
   \s{C_m}}{C_i} where $\pnt{q} = \ppnt{q}{i} \cup \ppnt{q}{m}$.  If
 $H_m \subseteq H_i$ holds, the proposition in question is
 proved. Otherwise, one keeps removing clauses from the order
 constraint of D-sequent $S$.

 Let $C_r$ be the clause of $H_m \setminus H_i$ with the largest
 index. Note that $m < r$ holds (and, hence, $i < m < r$) for the same
 reason $i < m$ does.  By applying Proposition~\ref{prop:subst} to
 remove clause $C_r$ from the order constraint of $S$ one produces a
 new D-sequent $S$ equal to \ods{q}{(H_i \cup H_m \cup H_r) \setminus
   \s{C_m,C_r}}{C_i} where $\pnt{q} = \ppnt{q}{i} \cup \ppnt{q}{m}
 \cup \ppnt{q}{r}$. Note that since $i < r$ and $m < r$, set $H_r$
 cannot contain $C_m$ or $C_r$.  If $(H_m \cup H_r) \setminus \s{C_r}$
 is a subset of $H_i$ the proposition in question is
 proved. Otherwise, one picks the clause of $(H_m \cup H_r) \setminus
 \s{C_r}$ with the largest index that is not in $H_i$ and removes it
 by applying the transformation of Proposition~\ref{prop:subst}.

 The procedure above goes one until one produces a D-sequent $S$ with
 order constraint $H_i \setminus \s{C_m}$. This procedure converges,
 since one always removes a clause with the largest index and so this
 clause cannot re-appear in the order constraint of $S$. Thus,
 eventually, in no more than $k-m$ steps, all the clauses that are not
 in $H_i \setminus \s{C_{m}}$ will be removed from the order
 constraint of $S$.  $\square$
\end{mmproof}

\end{document}